\documentclass[cernpreprint,texlive=2011,txfonts,USenglish]{atlasdoc}
\pdfoutput=1
\usepackage[biblatex=false]{atlaspackage}
\usepackage{atlasphysics}

\usepackage{graphicx}
\usepackage{multirow}
\usepackage{adjustbox}
\usepackage{rotating}

\DeclareGraphicsExtensions{.pdf}

\def\mG{\ensuremath{m_{G^*}}}
\def\coupling{\ensuremath{k/\overline{M}_{\mathrm{Pl}}}}
\def\mgg{\ensuremath{m_{\gamma\gamma}}}
\def\sigmaTbr{\ensuremath{\sigma\times BR(G^*\rightarrow \gamma\gamma)}}
\def\mpl{\ensuremath{M_{\mathrm{Pl}}}}
\def\mplRed{\ensuremath{\overline{M}_{\mathrm{Pl}}}}
\def\pythia{{\sc Pythia}}
\def\diphox{{\sc Diphox}}
\def\geant{{\sc Geant}}
\def\ETiso{\ensuremath{E_{\text{T}}^{\text{iso}}}}

\newcommand*{\NIMA}{Nucl.\ Instrum.\ Meth.\ A}

\AtlasTitle{Search for high-mass diphoton resonances in $pp$~collisions at $\sqrt{s}=8$~TeV with the ATLAS detector}
\AtlasAbstract{%
This article describes a search for high-mass resonances decaying to a pair of photons using a sample of $20.3$~fb$^{-1}$ of $pp$~collisions
at $\sqrt{s}=8$~TeV recorded with the ATLAS detector at the Large Hadron Collider. The data are found to be in agreement with the Standard Model
prediction, and limits are reported in the framework of the Randall-Sundrum model. This theory leads to the prediction of graviton states, the lightest
of which could be observed at the Large Hadron Collider.
A lower limit of $2.66$~($1.41$)~TeV at 95\% confidence level is set on the mass of the lightest graviton for couplings of $\coupling = 0.1$~($0.01$).
}

\author{The ATLAS Collaboration}

\PreprintIdNumber{CERN-PH-EP-2015-043}
\AtlasJournal{Phys.\ Rev.\ D.}
\date{\today}

\begin{document}

% placement of floats
\renewcommand{\topfraction}{0.99}
\renewcommand{\bottomfraction}{0.99}
\renewcommand{\textfraction}{0.001}

\section{Introduction}
One of the primary goals of the experiments at the Large Hadron Collider (LHC) is the search for new phenomena
which might manifest themselves in the high-energy regime made accessible for the first time by this machine. This article reports
on a search for new high-mass resonances in the diphoton channel using the full data set of proton-proton~($pp$) collisions recorded
with the ATLAS detector~\cite{Aad:2008zzm} at $\sqrt{s} = 8$~TeV in 2012.

The diphoton channel is important for searches for physics beyond the Standard Model~(SM) since new high-mass states decaying to two photons are
predicted in many extensions of the SM, and since searches in the diphoton channel benefit from a clean experimental signature: excellent mass
resolution and modest backgrounds. One popular example of theories predicting a new high-mass diphoton resonance is the Randall-Sundrum~(RS)
model~\cite{Randall:1999ee}. This model attempts to solve the so-called hierarchy problem of the~SM, i.e.\ the fact that the electroweak
scale is 16~orders of magnitude smaller than the Planck scale ($\mpl \approx 10^{16}$~TeV).
The hierarchy problem poses the question of the naturalness of the SM, and despite significant efforts over the past few decades, this problem has not been solved so far.
%The solution to this problem has been one of the major goals of particle
%physicists over the past few decades.
In the RS~model, the observed hierarchy of scales is accommodated assuming an extra spatial
dimension. In this paradigm, the fundamental strength of gravity is comparable to the strength of the electroweak interaction, but gravity appears
much weaker than the other interactions because it is diluted by the presence of the extra spatial dimension.

The RS~model postulates a ``warped'' five-dimensional spacetime configuration. In this configuration there are two 3-branes that are the boundaries of
the five-dimensional spacetime. The 3-branes can support ($3+1$)-dimensional field theories. The SM fields are located on one of the two 3-branes,
the so-called TeV~brane, while gravity originates from the other brane, called the Planck brane. Gravitons propagate in the five-dimensional
bulk. The fifth dimension is compactified with length~$r_{\mathrm{c}}$. The solution to Einstein's equations for this configuration is given by a spacetime metric
that couples the standard four dimensions to the extra dimension with an exponential ``warp factor''. In this configuration, a field with the fundamental
mass parameter~$m_0$ will appear to have the physical mass $m = e^{-k\pi r_{\mathrm{c}}} \, m_0$, where $k$ is the curvature scale of the extra dimension.
Scales of the order of 1~TeV are therefore generated from fundamental scales of the order of~\mpl\ if $kr_{\mathrm{c}} \approx 12$~\cite{Davoudiasl:1999jd}. The compactification of the extra
dimension gives rise to a Kaluza-Klein~(KK)~\cite{ModernKK} tower of graviton excitations~$G^*$ -- a set of particles on the TeV~brane with increasing mass. The phenomenology
can be described in terms of the mass of the lightest KK~graviton excitation~(\mG) and the dimensionless coupling to the SM~fields, \coupling,
where $\mplRed = \mpl/\sqrt{8\pi}$ is the reduced Planck scale. The masses and couplings of the individual graviton states are determined by the
scale $\Lambda_\pi = e^{-k\pi r_{\mathrm{c}}} \, \mplRed \approx {\cal O}(\text{TeV})$~\cite{Davoudiasl:1999jd}. Values of \coupling\ in the range $[0.01,0.1]$ are preferred from
theoretical arguments~\cite{Davoudiasl:1999jd}. The lightest graviton state is expected to be a fairly narrow resonance for $\coupling < 0.3$~\cite{Davoudiasl:1999jd}.
Its natural width varies like the square of~\coupling. For $\coupling = 0.1$, the natural width increases from $\sim 8$~GeV at $\mG = 800$~GeV to $\sim 30$~GeV
at $\mG = 2200$~GeV. In the first example, the natural width and the experimental mass resolution are similar, in the second example the natural width dominates the
total width~\footnote{The experimental mass resolution at $\mG = 2200$~GeV is only marginally (6\%) better than at $\mG = 800$~GeV.} of the mass peak.

Searches for high-mass diphoton resonances at the LHC have been reported by both the ATLAS and CMS collaborations, and the latest results are
summarized in Table~\ref{tab:Previous}. The most stringent experimental limits on RS~gravitons to date have been obtained at the LHC, using a combination of the
dielectron and dimuon channels. Limits from searches for narrow resonances in other channels are available in, for example, the dijet and $WW$~channels,
but these channels are not competitive for the specific case of RS~gravitons. The limits obtained using these channels are included in Table~\ref{tab:Previous}, along with those
from earlier searches at the Tevatron.

\begin{table*}[tb]
%\begin{center}
\caption{\label{tab:Previous}Overview of previous limits on the mass of the lightest RS~graviton state, expressed at 95\%~confidence level.
The notation $\ell^+ \ell^-$ refers to a combination of the dielectron and dimuon channels. The CDF and D0 experiments operated at the Tevatron
$p\overline{p}$~collider.}
\begin{adjustbox}{width=\textwidth}
 \begin{tabular}{lccc|lc}\hline \hline
Experiment & ~$\sqrt{s}$ [TeV]~ & ~Luminosity [fb$^{-1}$]~ & Channel        & Limit & ~Reference \\ \hline \hline
ATLAS      &    $7$           & $4.9$                 & $\gamma\gamma$ &  $2.06$~($1.00$)~TeV for \coupling \ = 0.1~(0.01) & \cite{Aad:2012cy} \\
CMS        &    $7$           & $2.2$                 & $\gamma\gamma$ &  $1.84$~($0.86$)~TeV for \coupling \ = 0.1~(0.01) & \cite{Chatrchyan:2011fq} \\
ATLAS      &    $8$           & $20.5$                & $\ell^+ \ell^-$ &  $2.68$~($1.25$)~TeV for \coupling \ = 0.1~(0.01) & \cite{Aad:2014cka} \\
CMS        &    $7$ and $8$   & $5.3$ and $4.1$       & $\ell^+ \ell^-$ &  $2.39$~TeV for \coupling \ = 0.1 & \cite{Chatrchyan:2012oaa} \\
CMS        &    $8$           & $4.0$                 & jet-jet         &  $1.45$~TeV for \coupling \ = 0.1 & \cite{Chatrchyan:2013qha} \\
ATLAS      &    $7$           & $4.7$                 & $W^+ W^-$       &  $1.23$~TeV for \coupling \ = 0.1 & \cite{Aad:2012nev} \\
CDF        &    $1.96$        & $5.7$ and $5.4$       & $e^+ e^-$ and $\gamma\gamma$ & $1.06$~TeV for \coupling \ = 0.1  & \cite{Aaltonen:2011xp} \\
D0         &    $1.96$        & $5.4$                 & $e^+ e^-$ and $\gamma\gamma$ & $1.05$~($0.56$)~TeV for \coupling \ = 0.1~(0.01)  & \cite{Abazov:2010xh} \\
\hline \hline
 \end{tabular}
\end{adjustbox}
%\end{center}
\end{table*}

\section{The ATLAS detector}

The ATLAS detector is described in detail in Ref.~\cite{Aad:2008zzm}. It consists of an inner detector (ID) surrounded by a solenoid that produces a
$2$~T magnetic field, electromagnetic~(EM) and hadronic calorimeters, and a muon spectrometer which employs air-core toroidal magnets. In the
pseudorapidity~\footnote{ATLAS uses a right-handed coordinate system with its origin at the nominal interaction point in the center of the detector
and the $z$-axis along the beam pipe. The $x$-axis points from the interaction point to the center of the LHC ring, and the $y$-axis points upward.
Cylindrical coordinates $(r,\phi)$ are used in the transverse plane, $\phi$~being the azimuthal angle around the beam pipe. The pseudorapidity is defined
in terms of the polar angle~$\theta$ as $\eta = - \ln \tan(\theta/2)$.}
region $|\eta|<2.5$, the ID provides measurements of the tracks originating from the primary $pp$~collision as well as tracks from
secondary vertices, permitting the efficient identification of photon conversions. The EM~calorimeter (ECAL) is a lead/liquid-argon (LAr) sampling calorimeter.
It is divided into a barrel section covering the region $|\eta|<1.475$ and two endcap sections covering $1.375<|\eta|<3.2$. The ECAL
is segmented longitudinally in shower depth into three layers for $|\eta|<2.5$ and two for $2.5<|\eta|<3.2$. Up to $|\eta|<2.4$, the first layer consists of highly
granular strips segmented in the $\eta$~direction for efficient candidate-by-candidate discrimination between single photon showers and two overlapping
showers originating from $\pi^0$~decay. The second layer, with a typical granularity of $\Delta\eta \times \Delta \phi = 0.025 \times 0.025$, collects most of the energy from photon showers. Significant energy
deposited in the third layer is an indication for leakage beyond the ECAL from a high-energy shower, and the measurements from the third layer
are used to correct for this effect.
A thin presampler layer, installed in front of the ECAL and covering the pseudorapidity interval $|\eta|<1.8$, is used to correct for energy loss before
the ECAL.
In the region $|\eta|<1.7$, the hadronic sampling calorimeter uses iron absorbers and plastic scintillator tiles as the active material.
In the region $1.5 < |\eta| < 4.9$, LAr is used as the active material, with copper or/and tungsten absorbers.
The muon spectrometer features three stations of precision chambers that allow for accurate measurements of the muon track curvature in the region $|\eta|<2.7$.
Fast detectors enable muon triggering in the range $|\eta|<2.4$.
A three-level trigger system is used to select the events to be recorded for subsequent offline analysis.

\section{Event reconstruction and selection}
\label{sec:EventSelection}

The events used in this analysis were recorded using a diphoton trigger. At the highest, software-based level of the trigger system, this trigger
requires at least one photon with transverse energy $E_{\mathrm T}^{\gamma} > 35$~GeV and at least one other photon with $E_{\mathrm T}^{\gamma} > 25$~GeV.
Both photons need to satisfy requirements on the shape of the energy deposit in the calorimeter. This trigger is nearly fully efficient for diphoton pairs that pass
the offline event selection.\\
In the offline analysis, only events that pass the standard ATLAS data quality requirements are considered.
Photons are reconstructed from fixed-size clusters of cells in the EM~calorimeter.
To ensure good diphoton mass resolution, photons are required to be located inside the precision region of the EM calorimeter ($|\eta|<1.37$ or $1.52<|\eta|<2.37$).
They are further required to satisfy $E_{\mathrm T}^{\gamma} > 50$~GeV as well as the {\em loose} identification requirements from Ref.~\cite{Aad:2014fxa} updated for 2012 run conditions.
The two highest-$E_{\mathrm T}^{\gamma}$ photon candidates that pass these requirements are retained for further analysis. Events are retained for the
main analysis if both candidates also satisfy the {\em tight} identification requirements~\cite{Aad:2014fxa} updated for the 2012 run conditions.
These requirements include upper limits on the energy leakage into the hadronic calorimeter and on the width of the shower in the first and second layers of the
EM~calorimeter. Events where at least one candidate fails the tight photon identification are retained for the study of backgrounds with misidentified (fake) photons in data control samples.\\
The invariant mass of the diphoton system is evaluated using the photon energies measured in the calorimeter as well as
the direction of the photon momenta determined from the positions of the photons in the calorimeter and the position of the primary vertex of the hard interaction.
Primary vertices are reconstructed from the tracks in the~ID and are required to be formed from at least three tracks.
The primary vertex of the hard interaction is identified using information on the directions of flight of the photons as determined using the longitudinal segmentation
of the ECAL (calorimeter pointing), the parameters of the beam spot and the properties of the tracks associated with each reconstructed vertex~\cite{Aad:2012tfa}.
This procedure ensures good diphoton mass resolution despite the presence of pileup.
Pileup refers to the additional $pp$~interactions that occur in the same bunch crossing as the $pp \rightarrow \gamma \gamma + X$ interaction or in the adjacent crossings.\\
A calorimetric isolation observable is defined for photon candidates. This observable provides further discrimination
between genuine photons and fake candidates from jets~($j$). It is used for two purposes in this analysis: to quantify the composition of the data sample in terms
of true and fake photon candidates, and to suppress the reducible background by requiring both photons to be isolated. The isolation energy~\ETiso\ is
defined as the sum of the transverse energies of cells inside clusters of variable size (also called ``topo-clusters''~\cite{Aad:2014bia}) inside a cone of
$\Delta R = \sqrt{(\Delta \eta)^2 + (\Delta \phi)^2} = 0.4$ around the photon candidate, excluding a $5\times 7$~grid of EM~cells in the
center of the cone.\\
The measurement of~\ETiso\ is corrected for the following two effects: i)~soft energy contributions from the underlying event and pileup that fall
inside the isolation cone and ii)~leakage of photon energy outside the central core, causing the isolation energy to grow as a function of photon~$E_{\mathrm T}^{\gamma}$.
The expected soft contributions are subtracted from~\ETiso\ on an event-by-event basis; the correction is based on a measurement of
the properties of low-$p_{\mathrm T}$ jets~\cite{Cacciari:2009dp} in the event. The average expected contributions to~\ETiso\ from leakage are subtracted based on simulations.
The fluctuations (from one photon shower to another) around the average expected leakage contribution to~\ETiso\ become more important
as $E_{\mathrm T}^{\gamma}$~increases. In particular, the tail at large~\ETiso\ becomes more pronounced and would cause significant inefficiencies at
large~$E_{\mathrm T}^{\gamma}$ if a fixed upper limit on~\ETiso\ were imposed in the event selection. To avoid inefficiencies,
a negative $E_{\mathrm T}^{\gamma}$-dependent offset is added to~\ETiso, and the requirement $\ETiso < 8$~GeV is imposed on both photon candidates after the
addition of the offset. In the following, the symbol~\ETiso\ refers to the isolation energy with the offset applied.
The $E_{\mathrm T}^{\gamma}$-dependence of the offset is chosen such that the efficiency of the \ETiso~requirement is
independent of~$E_{\mathrm T}^{\gamma}$ in simulated samples, and the size of the offset varies from zero at low~$E_{\mathrm T}^{\gamma}$ to $-5$~GeV at $E_{\mathrm T}^{\gamma} = 1$~TeV. The corresponding efficiency of the isolation requirement is $96$\%~per photon.
For the purpose of the determination of the background composition, a less stringent requirement of $\ETiso < 14$~GeV is used. Examples of the
\ETiso~distributions will be shown after the discussion of the backgrounds.

\section{Background estimate}

\begin{figure}[tb]
  \begin{center}
    \includegraphics[scale=0.4]{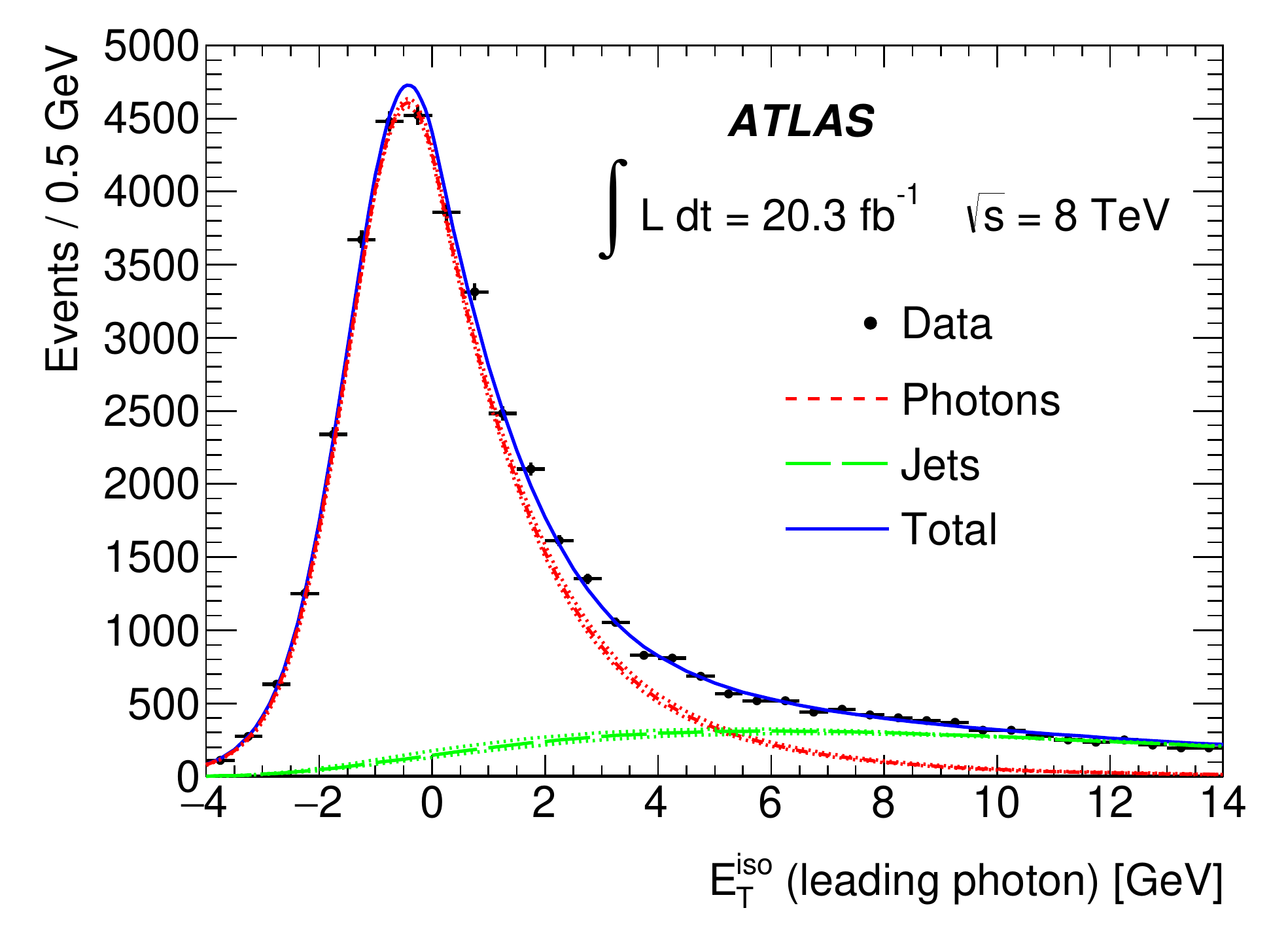}\\
    \includegraphics[scale=0.4]{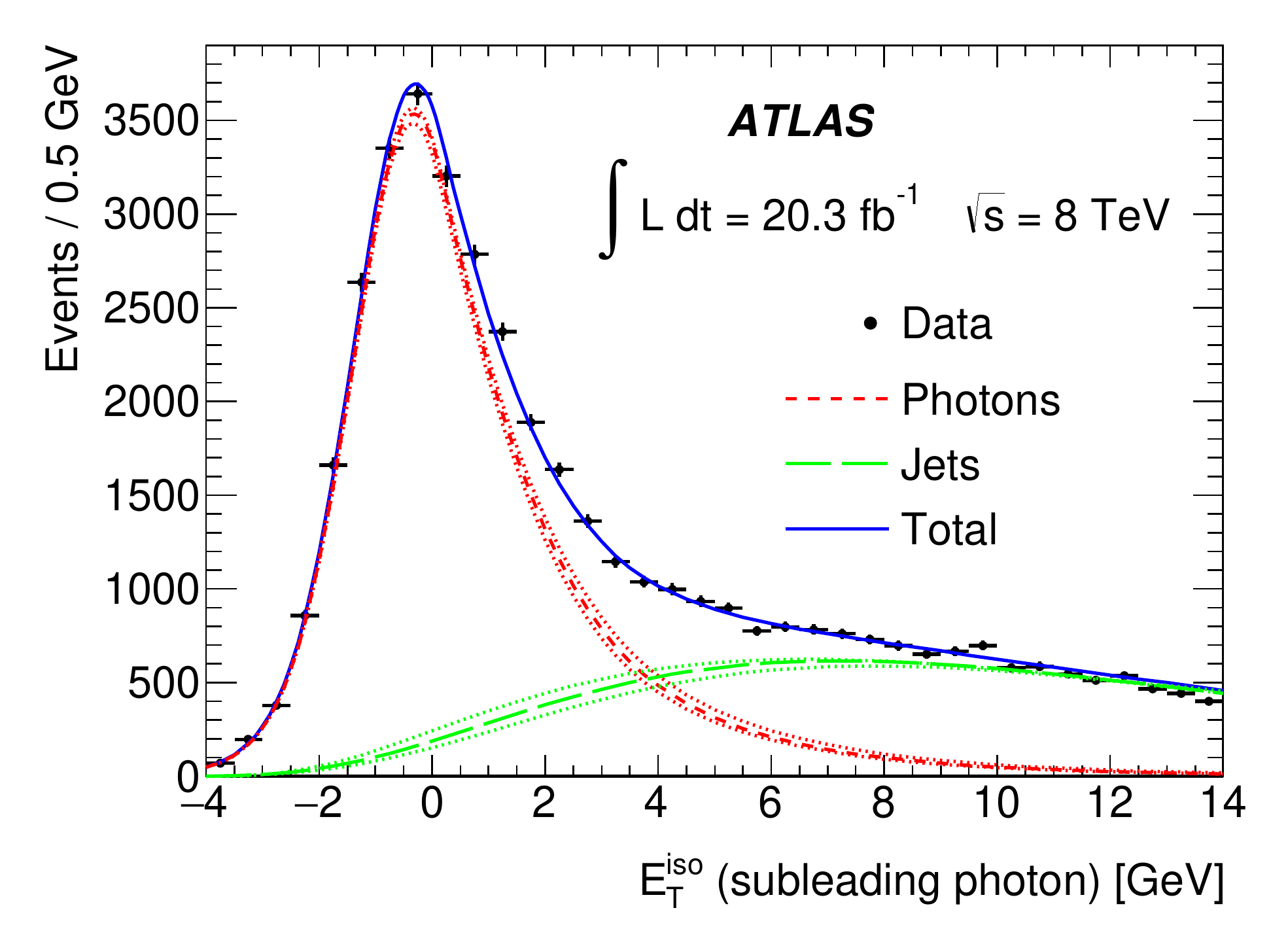}\\
    \includegraphics[scale=0.4]{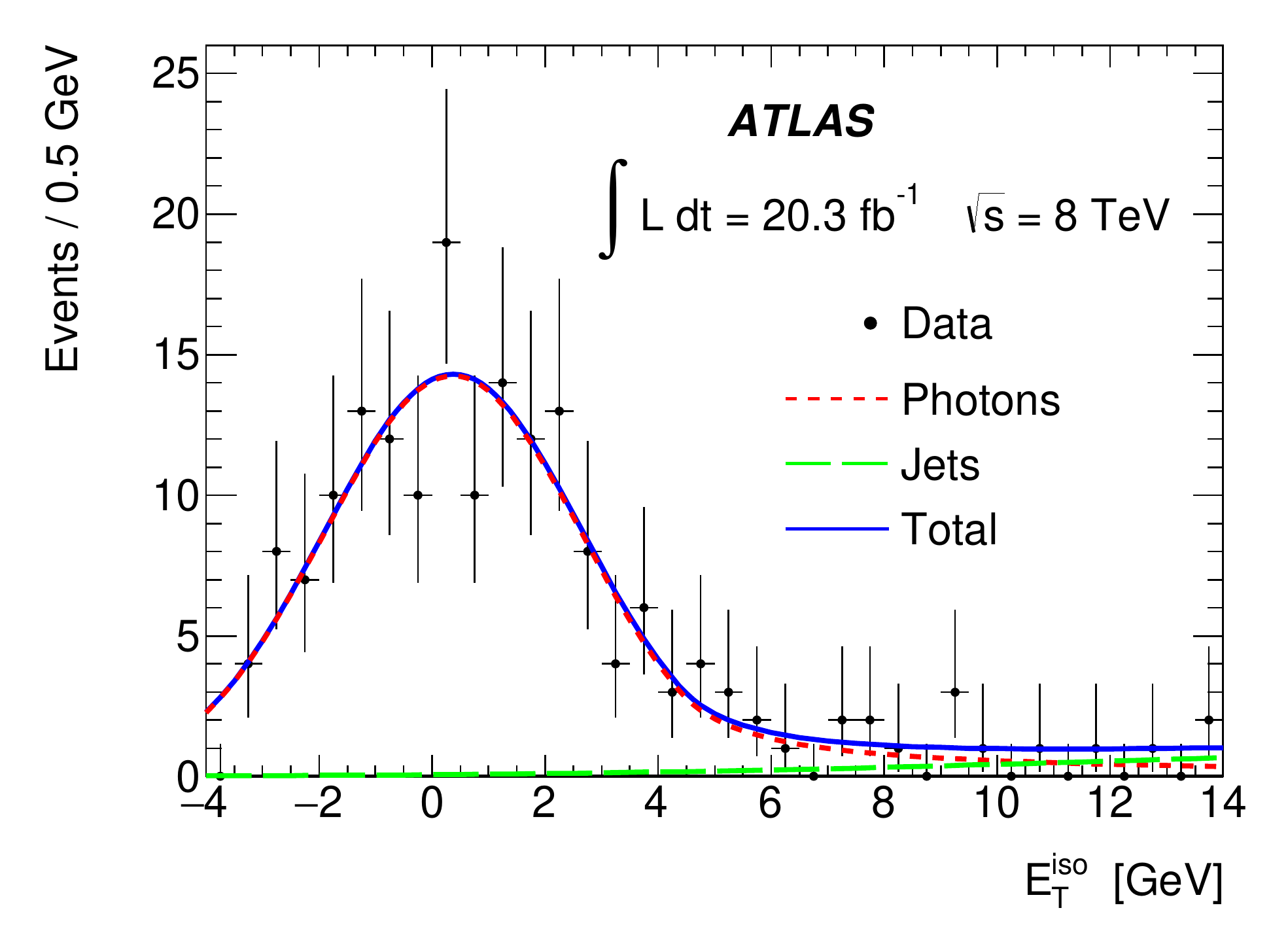}
    \caption{\label{fig:IsolationDiscussion}(Top) Distribution of the isolation energy~\ETiso\ for the leading photon candidate in events
    in the low-mass control region. %The full event selection described in the text, with the relaxed requirement $\ETiso < 14$~GeV, has been applied.
    (Middle) Corresponding distribution for the subleading photon. The black points represent the data. The blue
    line shows the result of the template fit to the data. The red and green lines show the templates for photon and jet~background respectively,
    normalized according to the fit result. The red and green bands represent the uncertainties in the photon and jet contributions due to the
    uncertainties in the template shapes. (Bottom) Distribution of~\ETiso\ for photon candidates with $E_{\mathrm T}^{\gamma} > 400$~GeV at any diphoton mass.
    The \ETiso\ values always include the $E_{\mathrm T}^{\gamma}$-dependent offset discussed in the text.}
  \end{center}
\end{figure}

An important contribution to the background arises from prompt $\gamma\gamma$~production via Standard Model processes. This contribution is irreducible and, as quantified below, represents the dominant source of background.
Another significant contribution is due to events in which one or both of the photon candidates arise from other objects, such as misidentified jets or electrons. This background component is dominated
by $\gamma + j$ and $j+j$~events with one or two jets reconstructed as photons. Backgrounds with electrons
misidentified as photons -- e.g.\ electron-positron events from Drell-Yan production, in $W/Z + \gamma$ or in $t\overline{t}$~events -- were verified
to be negligible after the event selection.\\
The irreducible background is estimated using Monte Carlo (MC) simulations normalized to the data in a control region defined below, and the reducible background is estimated using control samples in data.
A combination of the leading-order (LO) event generator \pythia~\cite{Sjostrand:2007gs} and the next-to-leading-order (NLO) parton-level simulation
\diphox~\cite{Binoth:1999qq} is used to simulate
%check the version of Pythia that was used to generate the SM background
the irreducible background. Monte Carlo samples are generated using \pythia~8.163 with the CTEQ6L1~\cite{Pumplin:2002vw} parton distribution
functions (PDFs) and the AU2 \pythia\ parameter tune~\cite{ATLAS:AU2}. The detector response is simulated using \geant~4~\cite{Aad:2010ah,Agostinelli:2002hh}, including pileup conditions similar to those observed in data.
The simulated events are reweighted such that the number of $pp$~interactions per bunch crossing has the same distribution as in data.
The NLO~correction and the contributions of the fragmentation process are evaluated using \diphox~1.3.2 with the CTEQ6.6M~\cite{Nadolsky:2008zw} PDF~set
and are used to scale the \pythia\ prediction by a factor that depends on the generated mass of the diphoton system.
The so-called box contribution $gg \rightarrow \gamma \gamma$ through a quark loop is included in both the \pythia\ and \diphox\ predictions.
From the point of view of power counting, this diagram is a next-to-next-to-leading-order (NNLO), i.e.\ ${\cal O}(\alpha_{EM}^2 \alpha_s^2)$, contribution. But given the large gluon luminosity at the LHC
compared to the quark-antiquark one, the contribution of the box diagram is comparable to that of the $q \overline{q} \rightarrow \gamma \gamma$ process, which corresponds to a LO, i.e.\ ${\cal O}(\alpha_{EM}^2)$, diagram.\\
To reduce the impact of the theory uncertainties in the absolute normalization of the production cross-section of the irreducible
background, the background estimates are scaled to the data at low mass. In the mass region $\mgg < 600$~GeV, where \mgg~denotes the measured invariant mass
of the diphoton system, the most sensitive searches for narrow resonances in the diphoton channel rely on background estimates that make use of low-mass and high-mass sidebands~\cite{Aad:2014ioa}. The search described in the present article focuses on higher diphoton masses where the sidebands, especially at high mass,
provide insufficient event yields. This search therefore uses a different approach to obtain the background prediction at high~\mgg. Specifically, the simulated
irreducible background is normalized to the data in a low-mass control region after subtraction of the reducible backgrounds. This control region is defined to coincide with the first $22$~bins in Fig.~\ref{fig:Mass}, i.e.\ as $179 < \mgg < 409$~GeV.
To determine the composition of the data sample in the low-mass control region in terms of irreducible and reducible backgrounds, a template fit to the \ETiso~distributions of
the leading and subleading photon candidates is performed. A similar technique was used in earlier analyses of low-mass diphoton events~\cite{Aad:2011mh,Aad:2011ww}.
The \ETiso~distributions, along with the result of the template fit, are shown in
Fig.~\ref{fig:IsolationDiscussion} (top and middle). The full event selection described in Sec.~\ref{sec:EventSelection}, with the relaxed requirement $\ETiso < 14$~GeV, has been applied.
The fit is performed simultaneously to the two distributions, and the normalizations for four
background components are allowed to float in the fit, namely the irreducible $\gamma\gamma$~component plus three reducible components:
$\gamma + j$ (subleading photon candidate is fake), $j + \gamma$ and $j+j$.
Templates for the \ETiso~distribution of true photons and of fake photons are determined from data. The template for fake photons is
obtained using a sample of photon candidates that pass the {\em non-tight} selection: photon candidates must satisfy the {\em loose} identification criteria
and fail to meet a subset of the criteria in
the {\em tight} selection. This subset has been chosen to include criteria that have been found to be only weakly correlated with~\ETiso.
The template for true photons is obtained from the sample of {\em tight} photon candidates after subtraction of the fake-photon template
normalized to the {\em tight} data sample at $\ETiso > 10$~GeV where fake photons dominate.\\  %% check the cut value
The fit is performed separately in five subsamples of the data, because the relative contributions of the four background components vary significantly
from one subsample to another. The five subsamples correspond to different combinations of pseudorapidities of the leading and subleading photon
candidates: both photons in the ``central'' region ($|\eta| < 1.37$), leading photon central and subleading photon in the ``endcap''
region ($1.52 < |\eta| < 2.37$), leading photon in endcap and subleading central, both photons in same endcap and both photons in endcaps of
opposite-sign pseudorapidity.
In Fig.~\ref{fig:IsolationDiscussion}, the data from the five subsamples, as well as the corresponding fit results are combined,
and only the sum of the photon and jet components over the four background components is shown.
To illustrate the broadening of the \ETiso~distribution at large~$E_{\mathrm T}^{\gamma}$ described above, Fig.~\ref{fig:IsolationDiscussion} also includes
the \ETiso~distribution for photons with $E_{\mathrm T}^{\gamma} > 400$~GeV.\\
The backgrounds in the search region, i.e.\ at masses higher than the low-mass control region, are predicted using estimates of the \mgg~shape of each background component and
the normalizations obtained in the low-mass control region described above. As discussed above, the shape of the irreducible component is obtained using simulation.
The shapes of the reducible components are obtained from data control samples. The control sample used to extract the shape
of the $\gamma + j$ background is selected in the same way as the signal sample, except that the subleading photon candidate has to satisfy the
{\em loose} identification criteria and fail to meet the {\em tight} criteria. The control samples for the $j + \gamma$ and $j+j$ backgrounds are defined using
the same approach. Since these data control samples contain relatively few events at high~\mgg, a fit to a smooth function of the form
$f(\mgg) = p_1 \times (\mgg)^{p_2 + p_3 \log(\mgg)}$, where the $p_i$ are free parameters, is used to extrapolate
the shapes of the reducible backgrounds to high~\mgg. This functional form describes well the shapes of the data in the control samples of
this analysis. It was also successful in describing the shapes of the background control samples in earlier searches in the dijet~\cite{Aaltonen:2008dn,Aad:2011fq},
dilepton~\cite{Collaboration:2011dca}, photon-jet~\cite{ATLAS:2011ai} and diphoton~\cite{Aad:2012cy} channels.\\
The shapes of the distribution of the isolation energy for genuine photons extracted from data (in bins of  $E_{\mathrm T}^{\gamma}$) are compared to
the shapes predicted using simulation. The impact of the small observed differences in the isolation distributions on the predicted \mgg~shape of
the irreducible background and on the signal efficiency is quantified and taken into account in the systematic uncertainties.\\

\begin{figure}[tb]
  \begin{center}
    \includegraphics[scale=0.5]{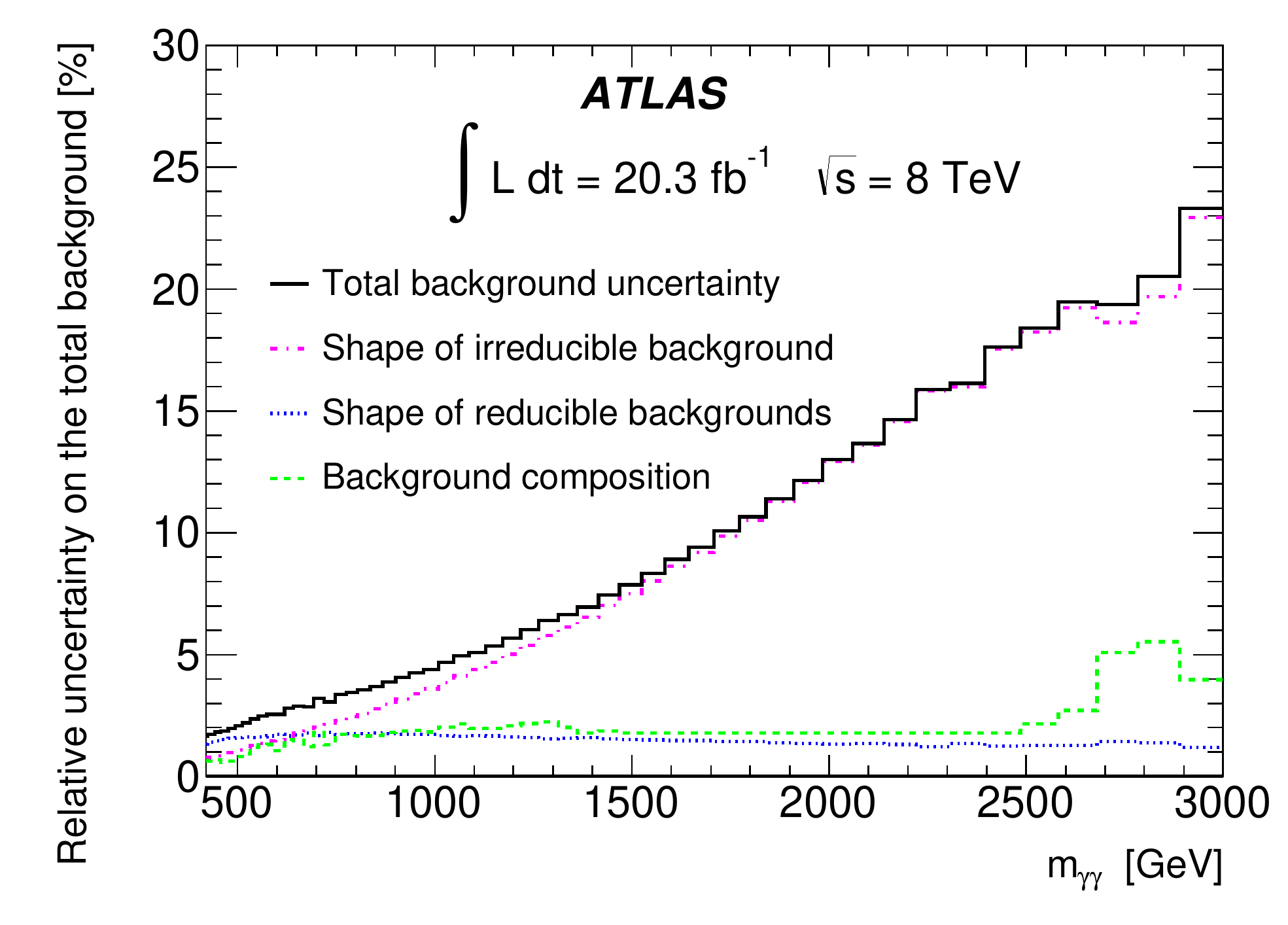}
    \caption{\label{fig:bkgndtotalsystematic} Uncertainty on the total background estimate as a function of~\mgg. The individual contributions
    due to the uncertainties in the irreducible and reducible components, as well as the contribution from the uncertainties in the normalization
    of the different background components (labeled ``background composition''), are also shown.}
  \end{center}
\end{figure}

The uncertainties in the prediction of the total background are shown in Fig.~\ref{fig:bkgndtotalsystematic} as a function of~\mgg. The uncertainties
in the shape of the \mgg~distribution of the irreducible background are dominated by the uncertainties in the PDF~set used for the \diphox\ simulation. The set of PDF variations provided with
the CTEQ6.6M PDF~set are used to propagate the PDF uncertainty to the shape of the irreducible background. The comparatively small difference between
the irreducible background shapes obtained with the CTEQ6.6M and the MSTW2008~\cite{Martin:2009iq} PDF~sets is added in quadrature to the uncertainty.
An additional uncertainty arises from  higher-order contributions that are not included in the \diphox\ generator. This uncertainty is evaluated by varying
in a correlated and anticorrelated way the renormalization scale and the factorization scales in the hard scattering process and in the photon fragmentation
function~\cite{Binoth:1999qq} by a factor two around their nominal value, which is set to the mass of the diphoton system.
A small uncertainty arises from residual imperfections in the simulation of the isolation energy~\ETiso.
The uncertainties in the shape of the reducible background arise from the finite size of the corresponding control samples, and from the extrapolation
of the background shapes from the control samples to the signal region. The latter are assessed by varying the definition of the {\em loose}
selection requirement that is used to define the control samples.
The uncertainties in the normalizations of the individual background components are dominated by the uncertainties in the template shapes that are
used in the fit to the isolation energy. They are assessed by varying the definition of the {\em non-tight} selection requirements that are used
to obtain the template shapes.
\\

\section{Signal response}

Simulated event samples of RS~graviton signals at different values of~\mG\ are used to study the response of the detector for signal events.
The samples are generated using \pythia~8.163 with the CTEQ6L1 PDF~set, the AU2 tune and the same detector and pileup simulation as for the
SM~$\gamma\gamma$ samples described above. The sources of uncertainty in the expected signal yield are
summarized in Table~\ref{tab:syst}. The systematic uncertainty is dominated by the uncertainty in the efficiency due to residual imperfections in the
simulation of the observables that are used in the {\em tight} photon identification criteria and in the simulation of the isolation energy. Smaller
contributions arise due to the limited size of the simulated samples and the determination of the trigger efficiency. The uncertainty
in the integrated luminosity is also propagated to the expected yield. Uncertainties in the RS~resonance shape due to our current
knowledge of the photon energy scale and resolution as well as imperfections in the simulation of the pileup conditions were verified to have a
negligible effect on the result.\\

\begin{table}[tb]
\begin{center}
\caption{\label{tab:syst}Summary of systematic uncertainties on the expected signal yield, excluding those on the production cross section and the acceptance.
The total systematic uncertainty is obtained by summing the individual contributions in quadrature.}
\begin{tabular}{l|c}\hline \hline
Source                                        & Uncertainty in signal yield [\%] \\ \hline \hline
Integrated luminosity                         & 2.8                     \\
MC statistics                                 & 1.0                     \\
Trigger efficiency                            & 1.0                     \\
Photon ID efficiency                          & 3.0                     \\
\multirow{2}{*}{Photon isolation efficiency}  & 0.3--2.1\\
                                              & ~~~~~~(for $\mG=500$--$3000$~GeV) \\\hline
Total                                         & $\approx$5                     \\ \hline \hline
\end{tabular}
\end{center}
\end{table}

\section{Results}

\begin{figure}[tb]
    \begin{center}
\includegraphics[scale=0.6]{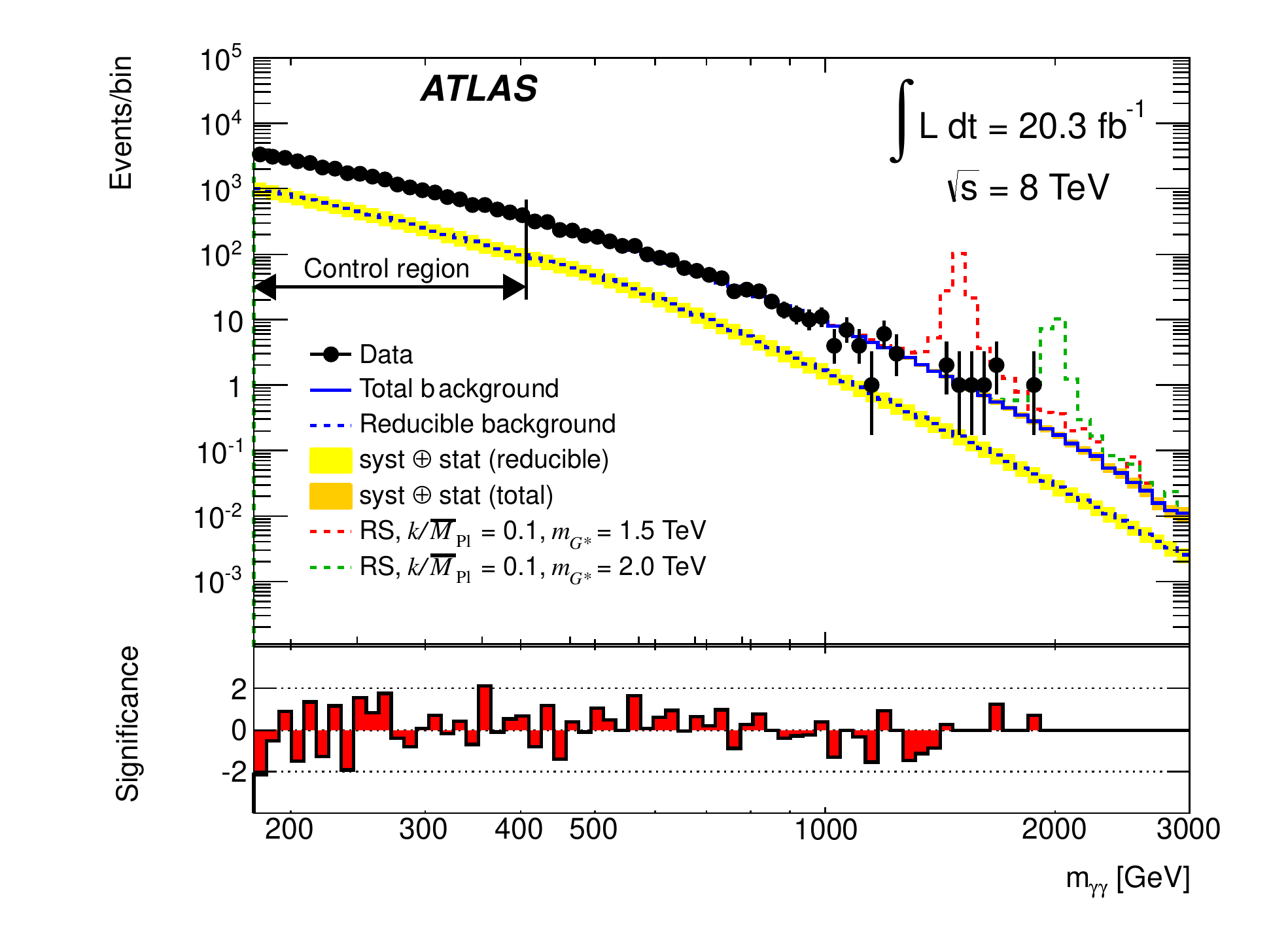}
\caption{\label{fig:Mass} Observed invariant mass distribution of the selected diphoton events (black dots; the vertical and horizontal axes on logarithmic scales).
To compensate for the rapid decrease of the spectrum, the bins are chosen to have constant logarithmic width.
Specifically, the ratio of the upper to lower bin boundary is equal to $1.038$ for all bins, and the first bin starts at $179$~GeV.
Superimposed are the SM~background prediction including irreducible and reducible components and two examples of signal predictions.
The low-mass control region is indicated by an arrow.
(Bottom) Bin-by-bin significance of the difference between data and background expectation.}
    \end{center}
\end{figure}

\begin{table*}
  \begin{center}
    \caption{\label{tab:bkgestimation} Number of events expected from the reducible and irreducible
    background components as well as the total background and observed
    number of events in different mass bins. Each bin boundary in the table corresponds to a
    bin boundary in Fig.~\ref{fig:Mass}, but the binning of the table is more coarse than in the figure.
    Since the combined statistical and systematic uncertainties are strongly anticorrelated between the two components,
    the uncertainty in the total background prediction
    is smaller than the sum in quadrature of the uncertainties of the individual components. }
    \begin{tabular}{ c |c c c | c} \hline
      \hline
      Mass window  & \multicolumn{3}{c|}{Background expectation (number of events)}  & Observed \\
      {[}GeV]        & Irreducible    & Reducible   & Total         & events   \\
      \hline
      \hline
      $[179,409]  $  & \multirow{2}{*}{~~23800 $\pm$ 2400~~} & \multirow{2}{*}{~~9100 $\mp$ 2400~~} &  \multirow{2}{*}{32866}  & \multirow{2}{*}{32866}\\
      Control region &                          &                     &         &            \\
      \hline
      $[409,513]  $ &      1070 $\pm$ 110   & 400 $\pm$ 100   &  1463 $\pm$ 27    &       1465 \\
      $[513,596]  $ &      369 $\pm$ 37     & 129 $\pm$ 34    &  498 $\pm$ 12     &       524  \\
      $[596,719]  $ &      240 $\pm$ 24     & 74 $\pm$ 20     &  314.4 $\pm$ 8.8      &       335  \\
      $[719,805]  $ &      75.8 $\pm$ 7.7   & 20.6 $\pm$ 5.5      &  96.4 $\pm$ 3.2       &       99   \\
      $[805,901]  $ &      46.6 $\pm$ 4.8   & 11.5 $\pm$ 3.2      &  58.1 $\pm$ 2.1       &       60   \\
      $[901,1009] $ &      28.2 $\pm$ 3.0       & 6.3 $\pm$ 1.8       &  34.5 $\pm$ 1.5       &       33   \\
      $[1009,1129]$ &      16.8 $\pm$ 1.9       & 3.4 $\pm$ 1.0       &  20.2 $\pm$ 1.0       &       15   \\
      $[1129,1217]$ &      6.92 $\pm$ 0.89        & 1.35 $\pm$ 0.46       &  8.27 $\pm$ 0.45        &       7    \\
      $[1217,1312]$ &      4.85 $\pm$ 0.73        & 0.88 $\pm$ 0.39       &  5.74 $\pm$ 0.36        &       3    \\
      $[1312,1415]$ &      3.11 $\pm$ 0.54        & 0.58 $\pm$ 0.28       &  3.69 $\pm$ 0.25        &       0    \\
      $[1415,1644]$ &      3.39 $\pm$ 0.59        & 0.61 $\pm$ 0.29       &  4.00 $\pm$ 0.32        &       5    \\
      $[1644,3000]$ &      2.12 $\pm$ 0.61        & 0.41 $\pm$ 0.22       &  2.52 $\pm$ 0.30        &       3    \\
      \hline
      \hline
    \end{tabular}
  \end{center}
\end{table*}

The observed diphoton mass spectrum is shown in Fig.~\ref{fig:Mass}, together with the background expectation and the predicted signal
for two examples of values of the RS~model parameters. The acceptance (i.e.\ the fraction of simulated events in which both photons
satisfy the requirements on $|\eta|$~and~$E_{\mathrm T}^{\gamma}$ described above) for the RS~scenario with $\mG = 2.0$~TeV and $\coupling = 0.1$ is $89.3$\%
and the selection efficiency for events within acceptance is $(55.9 \pm 2.6)$\%. The bottom inset in Fig.~\ref{fig:Mass} shows, bin-by-bin, the statistical significance of the
difference between the data and the expected background. This significance is calculated using the prescription in Ref.~\cite{Choudalakis:Significance}
and is plotted as positive (negative) for bins with an excess (deficit) of data. The observed event yields and expected backgrounds are also summarized in Table~\ref{tab:bkgestimation}.
No significant deviation of the data from the expected background
is observed. To further quantify the level of agreement, the {\sc Bump\-Hunter}~\cite{Choudalakis:2011qn} algorithm is used to search for any ``bump''
on top of the expected background, i.e.\ any excess of events localized in a relatively narrow mass region. Specifically, the {\sc Bump\-Hunter} algorithm
scans the mass distribution in windows of progressively increasing width in order to identify the window with the most significant excess of observed
events over the background anywhere in a predefined search region. In this analysis, the binned \mgg~spectrum from Fig.~\ref{fig:Mass} is used,
the search region is defined as $409 < \mgg < 3000$~GeV, and the window size is allowed to vary from one bin to half the number of bins in the
search region. The most significant excess is found in the region $494 < \mgg < 745$~GeV. The probability to observe anywhere in the search region an excess that is at least
as significant as the one observed in data and that arises from fluctuations in the background is~58\%. This confirms the absence of any
significant excess in the data.

\begin{figure}[tb]
    \begin{center}
\includegraphics[scale=0.6]{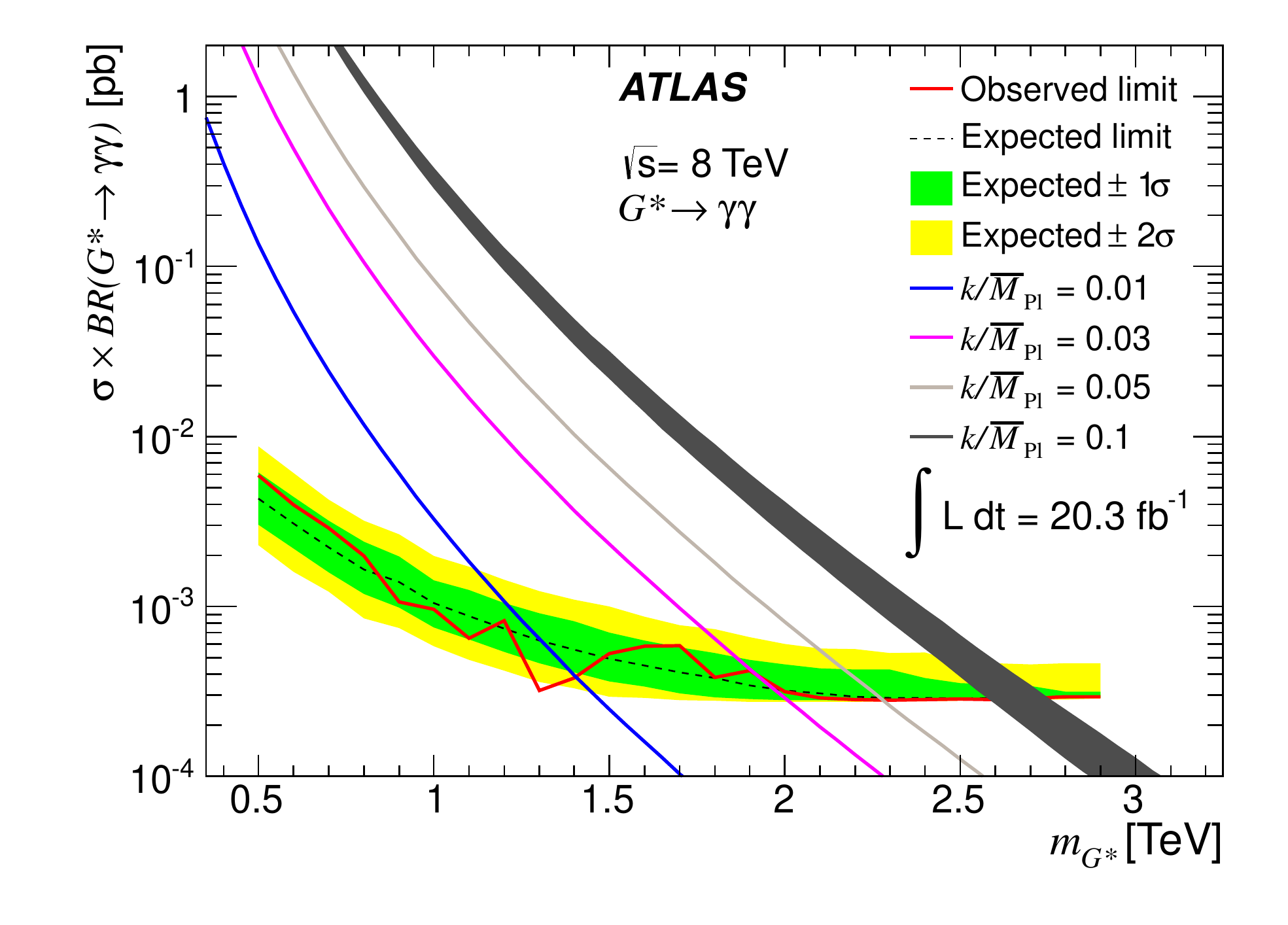}
\caption{\label{fig:limitXsec} Expected and observed upper limits on \sigmaTbr\ expressed at 95\%~CL, as a function of the
graviton mass. At large \mG, the $-1\sigma$ and $-2\sigma$ variations of the expected limit tend to be particularly close to the expected limit.
This is expected, since signals with high~\mG\ would appear in regions of~\mgg\ where the background expectation is small and the Poissonian
fluctuations around the mean expected background are highly asymmetric. The curves show the RS~model prediction for given values of \coupling\
as a function of \mG. They are obtained using the \pythia\ generator plus a $K$-factor to account for
NLO corrections (see text). The thickness of the theory curve for $\coupling = 0.1$ illustrates the PDF uncertainties expressed at 90\%~CL.}
    \end{center}
\end{figure}

\begin{figure}[tb]
    \begin{center}
\includegraphics[scale=0.6]{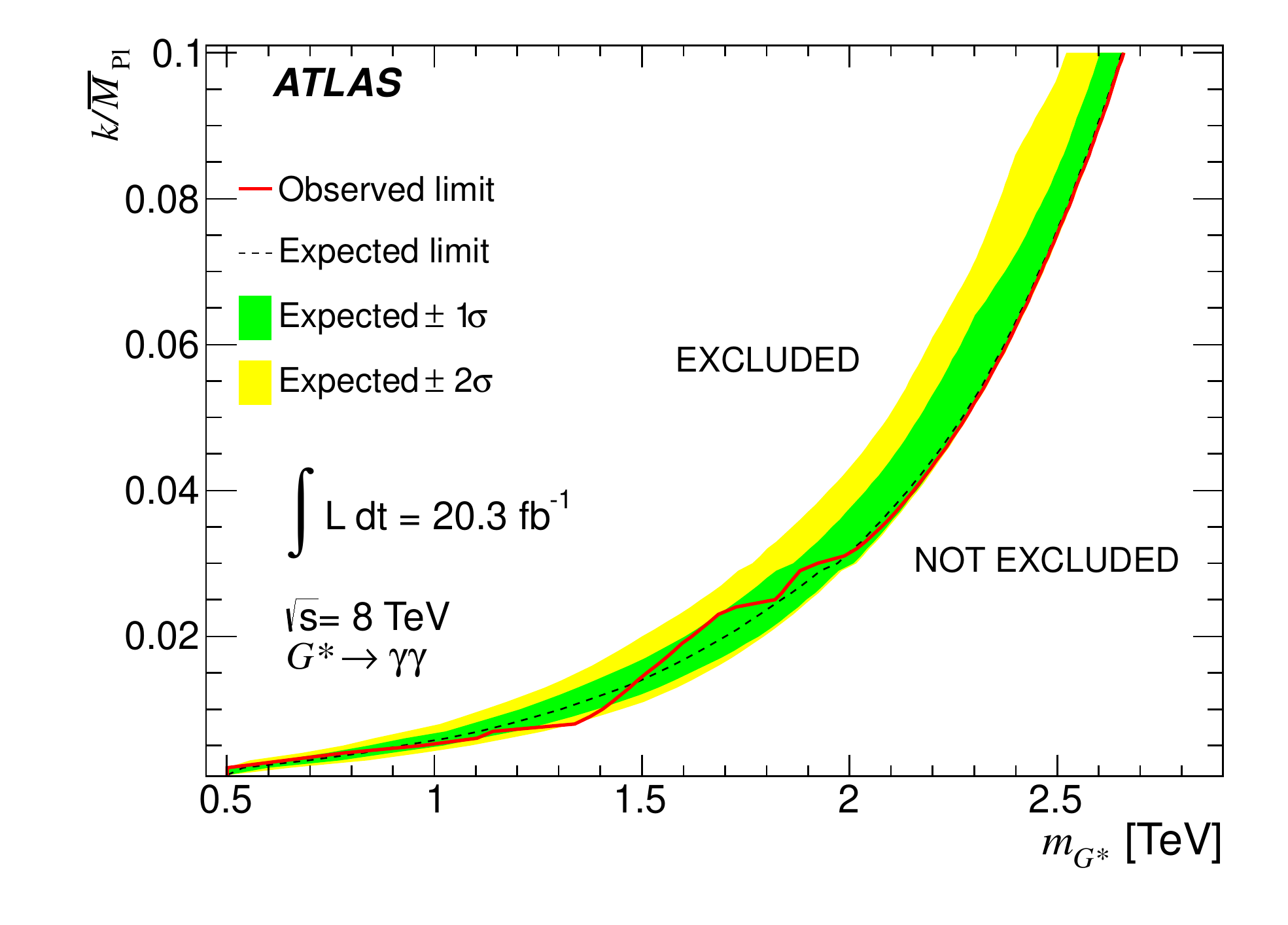}
\caption{\label{fig:limitCoupling} Expected and observed upper limits on \coupling\ expressed at 95\%~CL, as a function of the graviton mass.}
    \end{center}
\end{figure}

As no evidence for a signal is found, upper limits at 95\% confidence level~(CL) are set on the production cross section times branching ratio, \sigmaTbr, for the lightest
RS~graviton. A Bayesian approach is used to calculate the limits, and the corresponding computer program is implemented using the toolkit
from Refs.~\cite{Caldwell:2008fw,Beaujean:2011zz}. A likelihood function is used to describe the probability to obtain the \mgg~distribution
observed in data, given the background and signal predictions. The likelihood is defined as the product of the per-bin probabilities over all
bins in the search region ($\mgg > 409$~GeV) in Fig.~\ref{fig:Mass}. In each bin, the probability is computed from a Poisson distribution given
the observed number of events and the model prediction, where the model prediction is the sum of the expected numbers of background and signal
events. The expected signal yield depends on \mG\ and \sigmaTbr. Using Bayes' theorem, the likelihood is converted into a posterior probability.
A uniform positive prior in \sigmaTbr\ is assumed, and systematic uncertainties are incorporated using nuisance parameters with Gaussian priors.
For a given value of \mG, the posterior probability is reduced to one parameter of interest, \sigmaTbr, and the 95\%~CL limit on \sigmaTbr\ is
obtained via integration of the posterior probability.\\
The resulting limits on \sigmaTbr, as a function of the mass hypothesis (value of \mG), are shown in Fig.~\ref{fig:limitXsec}. The phenomenology of the RS~model can be described in terms of the parameters \mG\ and \coupling,
i.e.\ \sigmaTbr\ can be calculated~\cite{Han:1998sg,Allanach:2002gn} for a given pair of values
for these parameters. The results of this calculation are represented by the lines for given values of \coupling\ in Fig.~\ref{fig:limitXsec}.
They are obtained using the \pythia\ generator, which implements the calculations from Ref.~\cite{Allanach:2002gn}. A $K$-factor is applied
to these predictions to account for QCD~corrections at~NLO. The value of the $K$-factor varies between 1.6 and 1.9, depending on \mG\ and \coupling.
These results for the $K$-factor were provided by the authors of Refs.~\cite{Kumar:2008pk,Kumar:2009nn}, using updated calculations
at $\sqrt{s}=8$~TeV. Given these theory predictions for \sigmaTbr, the upper limits on \sigmaTbr\ from Fig.~\ref{fig:limitXsec} can be translated
into upper limits on \coupling. The results are shown, as a function of~\mG, in Fig.~\ref{fig:limitCoupling}. Alternatively, the upper limits
on \sigmaTbr\ can be translated into lower limits on the graviton mass. The results are shown in Table~\ref{tab:lim_sum}, for different values of~\coupling.
The limits presented in Fig.~\ref{fig:limitCoupling} and in Table~\ref{tab:lim_sum} do not take into account any theory uncertainties in the
prediction of \sigmaTbr\ for the RS~signal. The size of one of the dominant theory uncertainties (the PDF uncertainty) is included in
Fig.~\ref{fig:limitXsec}; uncertainties due to missing higher-order contributions in the calculation of the $K$-factor are estimated to be of comparable size.

\begin{table}[tb]
  \begin{center}
  \caption{\label{tab:lim_sum}Expected and observed lower limits at 95\%~CL on \mG\ for different values of \coupling.}
    \begin{tabular}{c||c|c|c|c|c||c}
      \hline \hline
               &\multicolumn{5}{c||}{Expected limit [TeV]}    &    Observed  \\
      \coupling & $-2\sigma$ & $-1\sigma$ & central & $+1\sigma$ & $+2\sigma$ &    limit [TeV]                           \\ \hline \hline
      0.010   &        1.45    &       1.38    &       1.30    &       1.20   &        1.11 &   1.41  \\
      0.020   &        1.80    &       1.78    &       1.70    &       1.60   &        1.49 &   1.62  \\
      0.030   &        2.02    &       2.01    &       1.96    &       1.86   &        1.77 &   1.92  \\
      0.040   &        2.17    &       2.16    &       2.14    &       2.04   &        1.94 &   2.15  \\
      0.050   &        2.29    &       2.28    &       2.27    &       2.17   &        2.08 &   2.28  \\
      0.060   &        2.38    &       2.38    &       2.37    &       2.27   &        2.19 &   2.38  \\
      0.070   &        2.47    &       2.46    &       2.45    &       2.37   &        2.27 &   2.46  \\
      0.080   &        2.54    &       2.54    &       2.53    &       2.46   &        2.38 &   2.54  \\
      0.090   &        2.61    &       2.60    &       2.60    &       2.53   &        2.45 &   2.60  \\
      0.100   &        2.67    &       2.66    &       2.66    &       2.60   &        2.51 &   2.66  \\ \hline \hline
    \end{tabular}
  \end{center}
\end{table}

\section{Conclusions}

In summary, a search for high-mass diphoton resonances has been performed using the full 2012 dataset collected by the ATLAS detector at the
LHC ($20.3$~fb$^{-1}$ of $pp$~collisions at $\sqrt{s}=8$~TeV). No significant excess over the expected background is observed,
and upper limits on the production cross section times branching fraction \sigmaTbr\ for narrow resonances are reported as
a function of the resonance mass.
High-mass diphoton resonances are predicted e.g.\ in models that postulate the existence of extra spatial dimensions in order to
address the hierarchy problem, and could be observed at the LHC. Limits on the mass of the lightest graviton state in the framework of the Randall-Sundrum model of extra dimensions are derived from the limits on \sigmaTbr. The only free parameters of this model  are the mass of the lightest graviton (\mG) and the coupling to the SM fields (\coupling),
i.e.\ \sigmaTbr\ can be calculated once the values of these two parameters are specified. The limits can therefore be expressed in terms
of \mG\ for a given value of \coupling.  A lower limit of $2.66$~($1.41$)~TeV at 95\% confidence level is obtained on the mass of the
lightest graviton for a coupling $\coupling = 0.1$~($0.01$).
The results reported in this article constitute a significant improvement over the results in Ref.~\cite{Aad:2012cy} obtained at $\sqrt{s}=7$~TeV.
The upper limits on \sigmaTbr\ are reduced by a factor of~$2.1$~($3.6$) for a resonance with a mass of $0.5$~TeV~($2.75$~TeV). Furthermore,
the new limits are expressed in terms of the production cross section at $\sqrt{s}=8$~TeV, compared to $\sqrt{s}=7$~TeV for the old limits,
and the production cross section for any new heavy particle is expected to be larger at $\sqrt{s}=8$~TeV.
The lower limits on the mass of the lightest RS~graviton are also significantly improved compared to the previous analysis.

\section*{Acknowledgements}
%\input{Acknowledgements}
% Acknowledgements for papers with collision data
% Version 23-Mar-2015
%--------------------------------------------------------------------------
We thank CERN for the very successful operation of the LHC, as well as the
support staff from our institutions without whom ATLAS could not be
operated efficiently.

We acknowledge the support of ANPCyT, Argentina; YerPhI, Armenia; ARC,
Australia; BMWFW and FWF, Austria; ANAS, Azerbaijan; SSTC, Belarus; CNPq and FAPESP,
Brazil; NSERC, NRC and CFI, Canada; CERN; CONICYT, Chile; CAS, MOST and NSFC,
China; COLCIENCIAS, Colombia; MSMT CR, MPO CR and VSC CR, Czech Republic;
DNRF, DNSRC and Lundbeck Foundation, Denmark; EPLANET, ERC and NSRF, European Union;
IN2P3-CNRS, CEA-DSM/IRFU, France; GNSF, Georgia; BMBF, DFG, HGF, MPG and AvH
Foundation, Germany; GSRT and NSRF, Greece; RGC, Hong Kong SAR, China; ISF, MINERVA, GIF, I-CORE and Benoziyo Center, Israel; INFN, Italy; MEXT and JSPS, Japan; CNRST, Morocco; FOM and NWO, Netherlands; BRF and RCN, Norway; MNiSW and NCN, Poland; GRICES and FCT, Portugal; MNE/IFA, Romania; MES of Russia and NRC KI, Russian Federation; JINR; MSTD,
Serbia; MSSR, Slovakia; ARRS and MIZ\v{S}, Slovenia; DST/NRF, South Africa;
MINECO, Spain; SRC and Wallenberg Foundation, Sweden; SER, SNSF and Cantons of
Bern and Geneva, Switzerland; NSC, Taiwan; TAEK, Turkey; STFC, the Royal
Society and Leverhulme Trust, United Kingdom; DOE and NSF, United States of
America.

The crucial computing support from all WLCG partners is acknowledged
gratefully, in particular from CERN and the ATLAS Tier-1 facilities at
TRIUMF (Canada), NDGF (Denmark, Norway, Sweden), CC-IN2P3 (France),
KIT/GridKA (Germany), INFN-CNAF (Italy), NL-T1 (Netherlands), PIC (Spain),
ASGC (Taiwan), RAL (UK) and BNL (USA) and in the Tier-2 facilities
worldwide.
%------------------------------------------------------------------------------

%%%%%%%%%%%%%%%%%%%%%%%%%%%%%%%%%Bibilo %%%%%%%%%%%%%%%%%%%%%%%%%%%%%%%%%%%%%%

\bibliographystyle{atlasBibStyleWoTitle}
%\bibliography{thebib}% Produces the bibliography via BibTeX.

\providecommand{\href}[2]{#2}\begingroup\raggedright\endgroup

%%%%%%%%%%%%%%%%%%%%%%%%%%End Biblio %%%%%%%%%%%%%%%%%%%%%%%%%%%%%%%%%%%%%%%%%%
%Author list
\clearpage
% ATLAS Collaboration author list
% Data extracted on 27-Mar-2015 for paper reference EXOT-2012-19
%\documentclass[11pt]{article}
%\usepackage{a4wide}\begin{document}
\begin{flushleft}
{\Large The ATLAS Collaboration}

\bigskip

G.~Aad$^{\rm 85}$,
B.~Abbott$^{\rm 113}$,
J.~Abdallah$^{\rm 152}$,
S.~Abdel~Khalek$^{\rm 117}$,
O.~Abdinov$^{\rm 11}$,
R.~Aben$^{\rm 107}$,
B.~Abi$^{\rm 114}$,
M.~Abolins$^{\rm 90}$,
O.S.~AbouZeid$^{\rm 159}$,
H.~Abramowicz$^{\rm 154}$
% ...
\bigskip
\\
% ...
$^{11}$ Institute of Physics, Azerbaijan Academy of Sciences, Baku, Azerbaijan\\
% ...
$^{85}$ CPPM, Aix-Marseille Universit{\'e} and CNRS/IN2P3, Marseille, France\\
% ...
$^{90}$ Department of Physics and Astronomy, Michigan State University, East Lansing MI, United States of America\\
% ...
$^{107}$ Nikhef National Institute for Subatomic Physics and University of Amsterdam, Amsterdam, Netherlands\\
% ...
$^{113}$ Homer L. Dodge Department of Physics and Astronomy, University of Oklahoma, Norman OK, United States of America\\
$^{114}$ Department of Physics, Oklahoma State University, Stillwater OK, United States of America\\
% ...
$^{117}$ LAL, Universit{\'e} Paris-Sud and CNRS/IN2P3, Orsay, France\\
% ...
$^{152}$ Institute of Physics, Academia Sinica, Taipei, Taiwan\\
% ...
$^{154}$ Raymond and Beverly Sackler School of Physics and Astronomy, Tel Aviv University, Tel Aviv, Israel\\
% ...
$^{159}$ Department of Physics, University of Toronto, Toronto ON, Canada\\
% ...
\end{flushleft}

%\end{document}
% Created with ./xml2latex.py 

\end{document}